\input harvmac
\overfullrule=0pt
\abovedisplayskip=12pt plus 3pt minus 3pt
\belowdisplayskip=12pt plus 3pt minus 3pt
\sequentialequations
%
\def\tilde{\widetilde}

\def\to{\rightarrow}
\def\tf{{\tilde\phi}}
\def\tb{{\tilde B}}
\def\td{{\tilde D}}
\font\zfont = cmss10 

\def\bigone{\hbox{1\kern -.23em {\rm l}}}
\def\ZZ{\hbox{\zfont Z\kern-.4emZ}}


\def\VAFAF{C. Vafa, {\it Evidence for F-theory}, hep-th/9602022.}
\def\VAFAETC{D. Morrison and C. Vafa, {\it Compactifications of
F-theory on Calabi-Yau Threefolds-I}, hep-th/9602114\semi
D. Morrison and C. Vafa, {\it Compactifications of F-Theory on
Calabi-Yau Threefolds-II}, hep-th/9603161\semi
M. Bershadsky, K. Intriligator, S. Kachru, D. Morrison, V. Sadov and
C. Vafa, {\it Geometric Singularities and Enhanced Gauge Symmetries},
hep-th/9605200.}
\def\TSEYTLIN{A.A. Tseytlin, {\it Selfduality of Born-Infeld Action 
and Dirichlet 3-brane of Type IIB Superstring Theory},
hep-th/9602064.} 
\def\WITPHASE{E. Witten, {\it Phase Transitions in M-Theory and
F-Theory}, hep-th/9603150.}
\def\WITSUPER{E. Witten, {\it Nonperturbative Superpotentials in String
Theory}, hep-th/9604030.}
\def\JATRAMA{D. Jatkar and S. Kalyan Rama, {\it F-Theory from
Dirichlet Three-branes}, hep-th/9606009.}
\def\FMSAG{S. Ferrara, R. Minasian and A. Sagnotti,
{\it Low-energy Analysis of M and F Theories on Calabi-Yau
Threefolds}, hep-th/9604097.}
\def\HORAVA{P. Horava, Nucl. Phys. {\bf B327} (1989) 461.}
\def\HORETC{A. Sagnotti, {\it Open Strings and Their Symmetry Groups},
in {\it Non-perturbative Quantum Field Theory}, Cargese 1987, eds. G.
Mack et. al. (Pergamon Press 1988)\semi
P. Horava, Phys. Lett. {\bf B231} (1989) 251\semi
J. Dai, R.G. Leigh and J. Polchinski, Mod. Phys. Lett. {\bf A4} (1989)
2073.}
\def\SEN{A. Sen, {\it F-theory and Orientifolds}, hep-th/9605150.}
\def\POLWIT{J. Polchinski and E. Witten, {\it Evidence for
Heterotic-Type I String Duality}, hep-th/9510169.}
\def\SEIWIT{N. Seiberg and E. Witten, hep-th/9408099, Nucl. Phys. {\bf
B431} (1994) 484.}
\def\BDS{T. Banks, M. Douglas and N. Seiberg, {\it Probing F-Theory
With Branes}, hep-th/9605199.}
\def\SEIBERG{N. Seiberg, {\it IR Dynamics on Branes and Space-time 
Geometry}, hep-th/9606017.}

{\nopagenumbers
\Title{\vtop{\hbox{hep-th/9606044}
\hbox{TIFR/TH/96-30}}}
{\centerline{F-theory at Constant Coupling}}
\centerline{Keshav Dasgupta\foot{E-mail: keshav@theory.tifr.res.in}
and Sunil Mukhi\foot{E-mail: mukhi@theory.tifr.res.in}}
\vskip 4pt
\centerline{\it Tata Institute of Fundamental Research,}
\centerline{\it Homi Bhabha Rd, Bombay 400 005, 
India}
\ \medskip
\centerline{ABSTRACT}

The subspace of the moduli space of F-theory on K3 over which the
coupling remains constant develops new branches at special values of
this coupling. These values correspond to fixed points under the
$SL(2,Z)$ duality group of the type IIB string. The branches contain
points where K3 degenerates to orbifolds of the four-torus by
$Z_3,Z_4$ and $Z_6$. A singularity analysis shows that exceptional
group symmetries appear on these branches, including pure $E_8\times
E_8$, although $SO(32)$ cannot be realised in this way. The orbifold
points can be mapped to a kind of non-perturbative generalization of a
IIB orientifold, and to M-theory orbifolds with non-trivial action on
2-brane wrapping modes.

\ \vfill 
\leftline{June 1996}
\eject} 
\ftno=0
\newsec{Introduction} 
Compactifications of the type IIB string in which the complex coupling
varies over a base are generically referred to as
``F-theory''. The simplest such construction corresponds to
elliptically fibred K3, equivalent to type IIB on $P^1$ with 24
7-branes\ref\vafaf{\VAFAF}. Other compactifications of F-theory, and
many miraculous properties of this class of models, have been
investigated in Refs.\vafaf\ref\vafaetc{\VAFAETC}%
\ref\tseytlin{\TSEYTLIN}\ref\witphase{\WITPHASE}%
\ref\witsuper{\WITSUPER}\ref\fmsag{\FMSAG}\ref\jatrama{\JATRAMA}.

Recently some new insights into the K3 compactification have been
obtained\ref\sen{\SEN} by considering a limit in which the coupling is
constant over the base. In this limit, the elliptic fibre always has a
constant, arbitrary complex-structure modulus $\tau$, but over 4
points of the base it degenerates to the singular fibre $T^2/Z_2$. The
entire K3 then degenerates to an elliptically fibred orbifold
$T^4/Z_2$. The advantage of considering this limit is that one can
explicitly map the problem to an orientifold of the type IIB
string. It turns out that physical effects which are nonperturbative
and hence invisible in the orientifold picture are nicely captured by
F-theory.

The limit considered in Ref.\sen\ has a 2 complex dimensional moduli
space, of which 1 complex dimension corresponds to the arbitrary value
of the coupling constant (there is also one real modulus, the size of
the base, which we will not refer to explicitly since it is always
present). Thus, there is a region of weak IIB coupling in this moduli
space. 

In this note, we observe that there are two other branches of the
F-theory moduli space on K3 where the coupling is constant over the
base. On these branches, the constant coupling is fixed to lie at one
of the fixed points of the moduli space of the elliptic fibre, namely
$\tau=i$ or $\tau=\exp(i\pi/3)$. The corresponding moduli spaces have
5 and 9 complex dimensions respectively. At special points within
these branches, the base becomes $T^2/Z_n$ and the whole K3 becomes
the orbifold $T^4/Z_n$, with the fibre becoming $T^2/Z_p$ over fixed
points of order $p$ (where $p$ divides $n$) on the base. Here
$n=4$ in the first branch, and $n=3,6$ in the second branch.

At these orbifold points, a singularity analysis predicts to various
enhanced-symmetry groups containing $E_6$, $E_7$ and $E_8$
factors. Also, following Ref.\sen\ one can map the theory to a kind of
orientifold\ref\horava{\HORAVA}\ref\horetc{\HORETC}\ of the type IIB
string. However, in this case the orientifold group includes
nonperturbative symmetries of the string theory.

\newsec{F-theory and K3 Orbifolds}
Elliptically fibred K3 surfaces can be defined by the family of
elliptic curves 
\eqn\ellip{
y^2 = x^3 + f(z) x + g(z) }
where $z$ is a coordinate on the $P^1$ base, and $f,g$ are polynomials of
degree 8,12 respectively in $z$. The modular parameter $\tau_f(z)$ of
the fibre is given in terms of the $j$-function by
\eqn\jfn{
j(\tau_f(z)) = {4(24 f(z))^3\over 4 f(z)^3 + 27 g(z)^2} }
The case of constant modulus corresponds to a situation where $f^3\sim
g^2$, in which case one can write
\eqn\fgphi{
\eqalign{ f(z) &= \alpha (\phi(z))^2\cr
g(z) &= (\phi(z))^3\cr}}
with $\phi(z)=\prod_{i=1}^4 (z-z_i)$ an arbitrary polynomial of degree
4 (whose overall coefficient can be scaled to 1). The constant
$\alpha$ determines the modular parameter by
\eqn\alphaj{
j(\tau_f) = {55296\alpha^3\over 4 \alpha^3 + 27}}

Thus we have obtained a subspace of the moduli space on which the
elliptic fibre has constant modulus $\tau_f$. Besides this modulus,
the subspace also has a complex parameter giving the location of one
of the zeroes of $\phi(z)$ (the other three locations can be fixed by
the $SL(2,C)$ invariance of $P^1$), and a real parameter corresponding to
the size of the base.

It has been argued in Ref.\sen\ that this subspace represents a K3
that has degenerated to $T^4/Z_2$, with the base $P^1$ having become
$T^2/Z_2$. The one free complex parameter in $\phi(z)$ represents the
complex structure of this base, which we call $\tau_b$, while the size
of the $P^1$ remains the real K\"ahler modulus of $T^2/Z_2$. In the
duality between F-theory on K3 and the heterotic string on
$T^2$\vafaf, this region can be mapped quite explicitly: the F-theory
side has gauge symmetry $SO(8)^4 \times (U(1))^4$, which means the
heterotic string has Wilson lines in the 9 and 10 directions breaking
$SO(32)$ or $E_8\times E_8$ to $SO(8)^4$, while the complex-structure
moduli $\tau_f$ and $\tau_b$ become the complex- and
K\"ahler-structure moduli $\tau$ and $\rho$ of the $T^2$ on which the
heterotic string is compactified. Finally, the size of the base in the
F-theory picture gets mapped to the heterotic string coupling.

It is also worth remarking that, in accordance with the analysis in
Ref.\ref\polwit{\POLWIT}, the heterotic string cannot develop any
enhanced symmetry in the presence of these Wilson lines, so there is
no prediction of any further enhanced gauge symmetries on the F-theory
side as we vary $\tau_b$ and $\tau_f$.

Two other branches of moduli space over which $\tau_f$ remains
constant emerge, very simply, in the limits $\alpha\to\infty$
and $\alpha\to 0$. The former limit corresponds (after a
rescaling) to taking $g(z)=0$, so that $j(\tau_f)= 13824$, from which
$\tau_f=i$. The latter limit gives instead $f(z)=0$, from which
$j(\tau_f)=0$ and $\tau_f = \exp(i\pi/3)$. In these limits, the
parametrization implied by Eq.\fgphi\ is no longer required, so the
theory develops a new branch in its moduli space. (It should be
stressed that these are not new branches of F-theory, but rather new
branches of the subspace of F-theory moduli space over which $\tau_f$
is constant.)

On the first branch, which we call branch (I) from now on, $f(z)$ is
an arbitrary polynomial of degree 8, while on the second (branch
(II)), $g(z)$ is an arbitrary polynomial of degree 12. After
subtracting an overall scaling and 3 $SL(2,C)$ parameters, one finds
that these branches of moduli space have complex dimension 5 and 9
respectively. Moreover, none of these parameters is the IIB coupling,
since that remains fixed on each branch. The size of the base of
course continues to be a real modulus.

Let us analyze the structure of the base on these
branches. On branch (I), Eqn.\ellip\ has the discriminant
\eqn\discrim{
\Delta(z) = \prod_{i=1}^8 (z-z_i)^3 }
where $z_i$ are the 8 zeroes of $f(z)$. Thus, generically, the base is
a $P^1$ with 8 singular points, at each of which there are 3 F-theory
7-branes, producing a deficit angle of $\pi/2$. These cannot be
thought of as orbifold singularities, since the deficit angle has to
be of the form ${n-1\over n}2\pi$ for a fixed point of order $n$.

Suppose we go to the special point in this moduli space where the 8
zeroes of $f$ have coalesced into 3 zeroes of order 3, 3 and 2. 
In this case, 
\eqn\discrimI{
\Delta(z) = (z-z_1)^9(z-z_2)^9(z-z_3)^6 }
The deficit angles are now $3\pi/2,3\pi/2$ and $\pi$ at the three
points. Thus we have two orbifold points of order 4, and one of order
2. This means the base has turned into $T^2/Z_4$, for which the
element of order 4 fixes 2 points and the element of order 2 fixes
another pair, which form a doublet under the $Z_4$ generator and count
as one point.

All the 5 moduli on this branch have to be fixed to achieve this, so
in this situation the base is completely fixed apart from its size. In
fact, the $Z_4$ quotient of a 2-torus is only defined if its modular
parameter is $i$. Thus we have an elliptically fibred K3 whose
base and fibre both have modulus $\tau_f=\tau_b=i$. Under a monodromy
around a fixed point of order 4 in the base, we have
\eqn\monod{
f \sim (z-z_1)^3 \to e^{6\pi i} f }
The equation defining the K3 is invariant under this only if we
also transform
\eqn\alsotrans{
\eqalign{
x&\to e^{3\pi i} x = -x\cr
y&\to e^{9\pi i\over 2} y = iy\cr}}
from which we conclude that the fibre above such a fixed point
has degenerated to $T^2/Z_4$. Altogether, this means that the K3 has
degenerated to $T^4/Z_4$. 

The monodromy in the fibre corresponds to the element 
\eqn\smatrix{
S=\pmatrix{0&-1\cr 1 &0\cr} }
which is of order 4 in $SL(2,Z)$. This monodromy operation transforms
the axion-dilaton pair $\tau = \tf + i e^{-\phi}$ by
\eqn\axdiltransf{
\tau\to -{1\over\tau} }
and the NS-NS and R-R 2-forms $(B, \tb)$ of the type IIB string 
by 
\eqn\bbtransf{
\eqalign{B &\to -\tb \cr
\tb & \to B\cr}}
The modulus $\tau_f=i$, which is the background value of the
axion-dilaton, is invariant under precisely this element, as expected.

Note that in the $T^4/Z_2$ orbifold limit studied in Ref.\sen, the
corresponding monodromy of the fibre was given by $diag(-1,-1)$, which
lies in the duality group $SL(2,Z)$ but acts as the identity in the
$PSL(2,Z)$ subgroup which is the nonperturbative part. In the present
case, the monodromy $S$ is nontrivial (and of order 2) in $PSL(2,Z)$,
and does not correspond to a perturbative symmetry of the type IIB
string.

Consider now the branch (II) in which $f(z)=0$. Here, we have
$\tau_f=\exp(i\pi/3)$ and there are generically 12 singular points on
the base, each with deficit angle $\pi/3$. This time, we can find two
interesting orbifold limits of the base. Consider first the case where
the 12 zeroes of $g(z)$ coalesce into three zeroes of order 5,4 and
3. Then,
\eqn\discrimII{
\Delta(z) = (z-z_1)^{10}(z-z_2)^8(z-z_3)^6 }
and the deficit angles are $5\pi/3, 4\pi/3$ and $\pi$. We can think of
these as fixed points of order 6,3 and 2 respectively. This is
precisely the structure of a $T^2/Z_6$ orbifold, whose modular
parameter is $\tau_b = \exp(i\pi/3)$. A monodromy about a fixed point
of order 6 transforms the fibre by the order-6 element of SL(2,Z):
\eqn\stmatrix{
ST = \pmatrix{0 &-1\cr 1 & 1} }
while around the point of order 3 we find $(ST)^2$, and around the
point of order 2, $(ST)^3 = S^2 =\, diag(-1,-1)$. Thus, this point is
just the limit where K3 has become $T^4/Z_6$.

Another interesting point in this space has the 12 zeroes coalescing
into three identical ones of order 4 each. For this case, it is easy
to see that
\eqn\discrimIII{
\Delta(z) = (z-z_1)^8(z-z_2)^8(z-z_3)^8 }
and we have a base $T^2/Z_3$, while the K3 has become $T^4/Z_3$. 

The only other points where the base can be thought of as an orbifold
are the ones for which $f$ develops 4 zeroes of order 2 each (in
branch (I)) or $g$ develops 4 zeroes of order 3 each (in branch
(II)). These are precisely the points where the base becomes $T^2/Z_2$
and where these branches join onto the branch of moduli space studied
in Ref.\sen.

At generic points on branch (I), where $f$ vanishes linearly, one can
check that the fibre above the singularity is again $T^2/Z_4$, while
at generic points of branch (II), where $g$ vanishes linearly, the
fibre is $T^2/Z_3$.

To summarise, our picture of the two branches is that on branch (I),
we have $\tau_f=i$ and a 5 complex dimensional moduli space, on which
the F-theory 7-branes can move in units of 3, while on branch (II),
$\tau_f=\exp(i\pi/3)$ and a 9 complex dimensional moduli space on
which the 7-branes move in units of 2.

\newsec{Enhanced Gauge Symmetries}
{}From the F-theory point of view, gauge symmetries arise from the
singularity type of the fibration. At a zero of the discriminant, one
uses Tate's algorithm to find the singularity type, and this has been
the subject of extensive study in recent months\vafaetc.

The cases of most interest to us are the ones where the base has
orbifold points of order 2,3,4 or 6. For order 2, near a zero at $z_1$
we have
\eqn\singtwo{
\eqalign{f(z)&\sim (z-z_1)^2\cr
g(z)&\sim (z-z_1)^3\cr
\Delta(z)&\sim (z-z_1)^6\cr} }
from which the singularity is of type $D_4\sim SO(8)$. This is the
case relevant to Ref.\sen. 

For order 3, we have
\eqn\singthree{
\eqalign{f(z)& =0\cr
g(z)&\sim (z-z_1)^4\cr
\Delta(z)&\sim (z-z_1)^8\cr} }
for which the singularity type is $E_6$.
For orbifold points of order 4, we get
\eqn\singfour{
\eqalign{f(z)&\sim (z-z_1)^3\cr
g(z)&=0 \cr
\Delta(z)&\sim (z-z_1)^9\cr} }
and the singularity type is $E_7$. Finally, for order 6 we get
\eqn\singsix{
\eqalign{f(z)&=0\cr
g(z)&\sim (z-z_1)^5\cr
\Delta(z)&\sim (z-z_1)^{10}\cr} }
for which the singularity type is $E_8$. It is rather remarkable that
the three exceptional groups arise precisely for the three allowed
types of orbifold points beyond $Z_2$.

Putting this together with the analysis of the previous section, we
find the following enhanced gauge symmetry groups for the three
orbifold limits of K3:
\eqn\allenhanced{
\eqalign{
T^4/Z_3:\qquad &E_6\times E_6\times E_6\cr
T^4/Z_4:\qquad &E_7\times E_7\times SO(8)\cr
T^4/Z_6:\qquad &E_8\times E_6\times SO(8)\cr}}

Note that these groups are all of rank 18, unlike the case of
$T^4/Z_2$ where the nonabelian part of the group has rank 16, and
indeed can only be $SO(8)^4$. Moreover, as explained earlier, one can
deform the above theories while preserving constancy of the
coupling. For example, for the branch with $\tau=\exp(i\pi/3)$, on
which the $Z_3$ and $Z_6$ points lie, one has a generic situation with
$f(z)=0$, $g(z)\sim (z-z_1)$ and $\Delta(z)\sim (z-z_1)^2$, for which
according to Tate's algorithm there is no singularity and hence we
expect the gauge group to be completely Abelian. On the
branch with $\tau=i$, we have at generic points $g(z)=0$, $f(z)\sim
(z-z_1)$ and $\Delta(z)\sim (z-z_1)^3$, for which the singularity is of
$A_1$ type and hence an SU(2) gauge symmetry appears there.

Finally, let us note that among the various gauge groups that can
appear, we have the possibility of realising pure $E_8\times E_8$. For
this, on branch (II) take
\eqn\eeight{
\eqalign{g(z)&= (z-z_1)^5 (z-z_2)^5 (z-z_3) (z-z_4)\cr
\Delta(z)&\sim (z-z_1)^{10} (z-z_2)^{10} (z-z_3)^2 (z-z_4)^2\cr}}
Then the two zeroes of $g(z)$ of order 5 give an $E_8$ factor each,
while as we have just seen, the simple zeroes give no
singularity\foot{In the first reference of \vafaetc, a somewhat
similar limiting configuration was suggested to give $E_8\times E_8$.}
Moreover, merging the zeroes at $z_3$ and $z_4$ will
produce an $SU(3)$ singularity there, so we will get $E_8\times
E_8\times SU(3)$. The only other allowed nonabelian gauge group of the
heterotic string on $T^2$ with unbroken $E_8\times E_8$ has the extra
factor $SU(2)\times SU(2)$, but this does not live on this branch of
F-theory moduli space. It cannot live on branch (I) either, since
there $E_8$ singularities cannot appear. Thus this vacuum evidently
cannot be realised by F-theory with constant coupling, and must lie
somewhere else in the full F-theory moduli space, where the coupling
varies. This is also true of the $SO(32)$ gauge group (and its
enhancements by $SU(2)\times SU(2)$ or $SU(3)$). Indeed, it is easy to
see that no $D_n$ group other than $D_4=SO(8)$ can arise in the moduli
space of constant coupling.

Let us briefly comment on the relationship of all this with the
heterotic string. First of all, the three $Z_n$ orbifold limits with
$n=3,4,6$ all give rank 18 gauge groups, hence in the heterotic string
they are special points in the Narain moduli space in which the
compactification torus is nontrivially mixed with the $E_8\times E_8$
torus. Hence the complex structure and K\"ahler structure moduli
$\tau_h$ and $\rho_h$ of the two-torus on which the heterotic string
is compactified should be fixed, and indeed they are, since one has
$\tau_h =\tau_f$ and $\rho_h=\tau_b$. As we have seen,
these are both fixed to be $i$ on branch (I) and $\exp(i\pi/3)$ on
branch (II). On the other hand, on branch II we can have the gauge
group $E_8\times E_8$, for which the heterotic string is just
compactified on the two-torus with no Wilson lines. This should admit
a decompactification limit to 10 dimensions, corresponding to
$\rho_h\to i \infty$. To find this, one has to identify the
appropriate modulus on branch (II) to take to infinity. One can only
roughly think of this as the parameter $\tau_b$, since now the base of
IIB is not a $T^2$ orbifold but a more general singular
$P^1$. However, it is clear that the configuration of 24 branes
grouped in units of 10,10,2 and 2 possesses precisely one modulus,
which has to be the relevant one for decompactification.

\newsec{Relation to Orientifolds}
In Ref.\sen, it was shown that F-theory on $T^4/Z_2$ maps to the
orientifold of type IIB on $T^2$ modded out by the $Z_2$ group $\{1,
\Omega (-1)^{F_L} I_{910}\}$ where $\Omega$ is reversal of world-sheet
orientation, $F_L$ is the spacetime fermion number of world-sheet 
left-movers, and $I_{910}$ is inversion of the 9th and 10th
dimensions. Since all the symmetries appearing in this $Z_2$ are
perturbatively realised, one can at least perturbatively define and
study this orientifold using conventional string theory
techniques. The vacuum contains 16 Dirichlet 7-branes, occurring 4 at
each fixed point of $Z_2$ on the base $T^2$. By a mechanism similar to
that studied in Ref.\polwit, moving the D-branes away from the fixed
points forces the coupling constant to vary over the base. 

When this happens, perturbative considerations in type IIB with
D-branes lead to an inconsistent description of the moduli space,
while the original F-theory formulation captures the nonperturbative
effects required to make this description consistent and correct. This
relationship between the perturbative IIB orientifold and F-theory
turns out to be identical\sen\ to that between the perturbative and
nonperturbative pictures of the moduli space in 4d N=2 supersymmetric
$SU(2)$ gauge theory with 4 hypermultiplets in the
fundamental\ref\seiwit{\SEIWIT}.

Something similar can be attempted on the branches of moduli space
that we have been studying, though a description through weakly
coupled D-branes will not be possible even in a limit. This is just as
well, since exceptional groups are not produced by D-branes at weak
coupling. So we will confine ourselves to a phenomenological
description of the situation.

Given the monodromies of the fibre at fixed points in the base, as
studied in Section 2, we can map F-theory on $T^4/Z_n$ to orientifolds
as follows. First, consider $n=4$. Over a fixed point of order 4, the
monodromy is given by the matrix $S$ in Eq.\smatrix. This acts on the
spacetime fields as described in Eqs.\axdiltransf\ and
\bbtransf. Denoting this action by $S$, and defining $R^{(4)}$ to be the
anticlockwise $\pi/2$ rotation of the base $T^2$ ($z\to iz$),
we can write this F-theory background as the ``generalised
orientifold'' type IIB on $T^2$ quotiented by
\eqn\zfour{
Z_4 = \{1, S.R^{(4)}, (S.R^{(4)})^2=\Omega(-1)^{F_L}I_{910}, 
(S.R^{(4)})^3\} }
As far as the action on the $T^2$ base is concerned, one has two fixed
points of order 4 and one doublet of order 2. Concentrating on the
order 2 point first, it is fixed under the element
$\Omega(-1)^{F_L}I_{910}$, hence this is a conventional orientifold
situation and should give rise, as usual, to 4 Dirichlet 7-branes at
the fixed point. This gives rise to the $SO(8)$ factor in the gauge
group (see Eq.\allenhanced).

We must now look for the $E_7\times E_7$ factor in the group. Each
$E_7$ must be associated with the ``twisted sector'' with respect to
the element $S.R^{(4)}$. This is precisely what we cannot describe by
conventional means. We can, however, indirectly argue some properties
of this sector. Suppose it is made of some new dynamical objects
(``E-branes''?). This configuration, which has 9 coincident F-theory
7-branes, can split in F-theory into a configuration localised at upto
three distinct points, each with 3 7-branes. This is because we are in
branch (I), where units of 3 7-branes must move around together. The
completely split case gives the gauge group $(SU(2))^3$, while the
case where one point splits from the other two gives $SO(8)\times
SU(2)$. Thus, any future understanding of such ``E-branes'', perhaps
as strongly coupled D-branes, will have to incorporate these
properties.

Similarly, on branch (II) at the $T^4/Z_6$ point, we can map the
F-theory to an orientifold of IIB on $T^2/Z_6$ where the $Z_6$ is
generated by the element $ST.R^{(6)}$ where $R^{(6)}$ is a rotation of
order 6. In this case, the point of order 2 again gets 4 conventional
Dirichlet 7-branes, giving $SO(8)$, while the point of order 3 gets an
$E_6$ gauge symmetry coming from 8 coincident F-theory
7-branes. Splitting of this point can now give $SO(8)\times U(1)$,
$SU(3)\times SU(3)$, $SU(3)\times U(1)^2$ and $U(1)^4$ depending on
whether the 8 branes split into (6,2), (4,4), (4,2,2) or
(2,2,2,2). Note that in this process the rank of the gauge group is
not preserved, unlike the behaviour of conventional D-branes. It would
be very interesting to understand better, if this is at all possible
when the coupling is of order 1, the behaviour of these unusual
dynamical objects. 

The generalized orientifolds described above can also be mapped to
other orientifolds of IIB and M-theory by T-duality. Suppose first
that we T-dualise in the directions 9 and 10. We concentrate on the
$T^4/Z_4$ case for definiteness. The order-2 element
$\Omega(-1)^{F_L}$ of the $Z_4$ group gets mapped to $\Omega$,
following the observations in Ref.\sen\ for the $T^4/Z_2$ case. As for
the order 4 element, it must map to something which is a square root
of $\Omega$. This can be deduced from the action of $\Omega$ on
spacetime fields. We have
\eqn\actionof{
\eqalign{\Omega:\qquad B&\to -B\cr
\tf&\to -\tf \cr 
\td &\to - \td \cr}}
where $B$ is the NS-NS 2-form and ${\tilde\phi}$, ${\tilde D}$ are the
R-R 0-form and 4-form. Under compactification to 8 dimensions these
fields give:
\eqn\fields{
\eqalign{
\tf &\to \tf \cr
B_{MN} &\to B_{\mu\nu},~ B_{\mu 9},~ B_{\mu 10},~
B_{910}\cr
\td_{MNPQ} &\to \td_{\mu\nu\lambda 9},~\td_{\mu\nu 910} \cr}}
Here we have not listed $\td_{\mu\nu\lambda 10}$ and
$\td_{\mu\nu\lambda\rho}$ since, by self-duality of the original $\td$
in 10 dimensions, these components are related to the ones listed
above.

Thus we have a pair of scalars, a pair of 1-forms, a
pair of 2-forms and a single 3-form. Now the action of the square
root of $\Omega$ must preserve 8-dimensional Lorentz invariance. Thus
scalars must be mapped to linear combinations of scalars, 1-forms
should be mapped to 1-forms and so on. This uniquely determines the
square root of $\Omega$ to be the transformation
\eqn\squarert{
{\tf\to B_{910}\atop B_{910}\to -\tf}\qquad 
{B_{\mu 9}\to B_{\mu 10}\atop B_{\mu 10}\to -B_{\mu 9}} 
\qquad
{B_{\mu\nu}\to \td_{\mu\nu 910}\atop \td_{\mu\nu 910} \to -B_{\mu\nu}}}
along with a duality on the 3-form $\td_{\mu\nu\lambda}$ (which, as
usual, means we take the Poincar\'e dual of its field strength). The
square of this duality in 8 dimensions is (-1), and clearly the
transformations in the equation above square to (-1) as well. 

Calling this combined transformation $U$, we see that it is part of
the U-duality group of type IIB in 8 dimensions. Thus after T-duality
we find that this orientifold is type IIB on $T^2$ modded out by the
$Z_4$ group given by $\{1,U,U^2=\Omega, U^3\}$. So even after
T-duality, the $Z_4$ group still depends on compactifying the theory
to 8d, unlike the $Z_2$ case where one found that all reference to the
compactified dimensions disappeared after T-dualising.

Now suppose instead that we had T-dualised only in one direction. In
the $Z_2$ case, this would take us to type IIA on the dualised
orientifold. However, in the present case two things change: on the
one hand, the presence of an $S$-duality transformation in the
orbifolding means that the T-dualised theory must be thought of as
M-theory rather than type IIA. In M-theory the $S$-duality is realised
as an interchange of the 11th and 10th directions. Second, the
rotation of order 4 on the compactification torus becomes, after one
T-duality, a rotation between momentum modes of IIA in the 9-direction 
and winding modes in the 10-direction. In M-theory language, this
orbifold group therefore interchanges modes of the 2-brane that wind
on a 2-torus, with modes that wind on a circle and propagate on the
other circle. Such an orbifold may not be out of reach of analysis,
and should provide further insight into F-theory and its relation to
M-theory, but we will not pursue it here.

\newsec{Discussion and Conclusions}
We have shown that the simplest compactification of F-theory, on K3,
has various regions of moduli space where the F-theory coupling
remains constant over the base. One branch of this region has been
utilised by Sen\sen\ to map to a problem involving conventional
D-branes, and to show that the moduli space of this problem is
governed by the Seiberg-Witten analysis of the moduli space of certain
4-dimensional N=2 supersymmetric gauge theories. These gauge theories,
in turn, have recently been interpreted in terms of the worldbrane
actions for Dirichlet 3-branes used as probes\ref\bds{\BDS}\foot{This
reference also suggested the possible relevance of $Z_n$ orbifolds to
the problem of getting exceptional groups.}. Even more recently it
has been argued that this framework is a powerful tool for analysing
the dynamics of supersymmetric gauge theories\ref\seiberg{\SEIBERG}.

Clearly, one should ask which 4d gauge theory, if any, relates to our
case. However, precisely because the orientifold related to our theory
is not of conventional type, one cannot in any obvious way introduce
3-branes as probes. A related fact is that the IIB coupling (which
would turn into the gauge coupling on the brane worldvolume), cannot
be taken small in the region that we study.

If a gauge theory description of this moduli space nevertheless exists
(and realises the $E$-series gauge symmetries of F-theory as global
symmetries) then this would be evidence that the concept of 3-branes
as probes could make sense beyond the context of conventional
Dirichlet 3-branes. Alternatively, it might suggest that strongly
coupled D-branes exhibit unusual behaviour, including the possibility
of producing exceptional gauge groups.

Finally, it has been speculated\witphase\ that at $\tau=i$ or
$\tau=\exp(i\pi/3)$, the type IIB string might exhibit some new
properties, analogous to tensionless strings. Since these points
coincide with the region of F-theory moduli space discussed in this
note, one might hope to see these new properties by examining the dual
heterotic description.
\bigskip
\noindent{\bf Acknowledgements} 
\ \smallskip
We are grateful to Rajesh Gopakumar, Kirti Joshi and Gautam Mandal for
helpful discussions.

\listrefs   
\bye